\documentclass[letterpaper, 10 pt, conference]{ieeeconf}
\usepackage{color}
\usepackage[english]{babel}
\usepackage[utf8]{inputenc}
\usepackage{amsmath}
\usepackage{graphicx}

\newcommand{\row}[1]%
{\mathord{\buildrel{\lower3pt%
\hbox{$\scriptscriptstyle\rightarrow$}}\over #1}}

\newcommand{\ket}[1]{\bigl| #1 \bigr\rangle}
\newcommand{\expect}[1]{\left\langle #1 \right\rangle}

\IEEEoverridecommandlockouts                              % This command is only
\title{\LARGE \bf
 Maximizing the encoded information via freezing the estimated parameters of a pulsed driven qubit }

\author{N. Metwally$^{a,b}$ and S. S. Hassan$^{a}$
\vspace{0.1cm}\\ % <-this % stops a space\\
$^{a}$Department of Mathematics, College of Science, University Of Bahrain, P. O. Box 32038, Bahrain\\
$^{b}$Department of Mathematics, Faculty of Science, Aswan
University, Aswan, Egypt.
\thanks{N. Metwally  {\tt\small Nmetwally@uob.edu.bh}}%
}

\begin{document}

\maketitle

% % \thispagestyle{empty}
% % \pagestyle{empty}

% \thispagestyle{empty}
% \pagestyle{empty}

%%%%%%%%%%%%%%%%%%%%%%%%%%%%%%%%%%%%%%%%%%%%%%%%%%%%%%%%%%%%%%%%%%%%%%%%%%%%%%%%
\begin{abstract}

We  use a rectangular pulse to freeze the possibility of
estimating  the  coherent parameters ($\theta,\phi$) of a single
qubit and the encoded information. It is shown that, as the
possibility of estimating the parameters increases, the amount of
encoded information decreases. The pulse strength and the detuning
between the qubit and the pulse have a different effect on the
estimation degree and the encoded information. We show that if the
weight parameter, $\theta$ is estimated, the encoded information
depends on the initial state settings. Meanwhile, the encoded
information doesn't depend on the estimated phase
parameter,$\phi$. These results may be useful in the context of
quantum cryptography, teleportation and secure communication.

\end{abstract}

%%%%%%%%%%%%%%%%%%%%%%%%%%%%%%%%%%%%%%%%%%%%%%%%%%%%%%%%%%%%%%%%%%%%%%%%%%%%%%%%
\section{INTRODUCTION}
It is well known that quantum information tasks. e. g. quantum
cryptography, quantum encoding \cite{bin,Metwally2011} and quantum
computation \cite{Nilsen}. require pure states to be implemented
with high efficiencies. However, decoherence is an inevitable
process due to the interaction with the surroundings. There are
different techniques that have been introduced to protect these
states' decoherence. Among of these methods are  quantum
purification \cite{bin96}, weak measurement \cite{Guo}, and
quantum filtering \cite{Yash}.

Recently, it was shown that  different pulse shapes  can keep the
quantum correlation survival and  consequently, the phenomena of
the longed lived entanglements is depicted \cite{Shoukry}. Very
recently,  Metwally and Hassan \cite{Shoukry1} investigated  the
initial parameters which describe the pulsed  driven state that
maximize/ minimize the Fisher information which  contained in the
driven state. However, in our previous work, we showed that  the
possibility of estimating these parameters is very small during
the pulse duration and for some cases it is frozen. This means
that, one  may estimate the these parameters within a certain
constant value    during the pulsed time and consequently, if this
state is captured by any Eavesdropper, may he/she get a minimum
information or nothing at all. Theses observations motivated  us
to investigate the possibility of freezing \cite{Metwally2018} the
pulsed qubits from a sender to a receiver by using the rectangular
pulse.

 The paper
is organized as following. In Sec.(2), we describe the initial
system and its driving by the  rectangular pulse. In Sec.(3), we
evaluate the encoded information of the  driven qubit. Finally, we
summarize our result in Sec.(4).

\section{The suggested Model }

Here, we consider a single qubit taken as 2-level atomic
transition of frequency $\omega_q$ and driven by a short laser
pulse of arbitrary shape and of circular frequency $\omega_c$ in
the absence of any dissipation process. The quantized Hamiltonian
of the system (in units of $\hbar=1$) in the dipole and rotating
wave approximation  and in a rotating frame of $\omega_c$ is given
by\cite{Shoukry1},

\begin{equation}\label{Ham}
\hat{H}=\Delta\hat{\sigma_z}+\frac{\Omega(t)}{2}(\hat{\sigma_{+}}+\hat{\sigma_{-}})
\end{equation}
where,  the spin-$\frac{1}{2}$ operators $\hat{S}_{\pm,z}$  obey
the $Su(2)$ algebra,
\begin{equation}\label{Com}
[\hat{\sigma_{+}}, \hat{\sigma_{-}}]=2\hat{\sigma_{z}},\quad
[\hat{\sigma_{z}},\hat{\sigma_{\pm}}]=\pm\hat{\sigma_{\pm}}
\end{equation}
and $\Delta=\omega_q-\omega_c$ is the atomic detuning and
$\Omega(t)=\Omega_o f(t)$, is the real laser Rabi frequency with
$f(t)$ is the pulse shape. Heisenberg equation of motion for the
spin  operators
$\hat{\sigma_x}=\frac{1}{2}(\hat{\sigma_+}+\hat{\sigma_{-}})$,
$\hat{\sigma_y}=\frac{1}{2i}(\hat{\sigma_+}-\hat{\sigma_{-}})$ and
$\hat{\sigma_z}$ according to (1), (2) are of the form,
\begin{eqnarray}
\hat{\sigma'}_x&=&-\Delta\hat{\sigma_y}
\nonumber\\
\hat{\sigma'_y}&=&\Delta\hat{\sigma_x}-\Omega(t)\hat{\sigma}_z \nonumber\\
\hat{\sigma'_z}&=&=\Omega(t)\hat{\sigma_y}
\end{eqnarray}
In the case of a rectangular pule of a short duration $T$(much
smaller than the life time of the qubit), we have
$\Omega(t)=\Omega_0; f(t)=1, t\in[0,T]$ and zero otherwise. In
this case, the exact solution of  the average Bloch vector
components $s_{x,y,z}(t)=\expect{{\hat\sigma}_{x,y,z}(t)}$ is the
matrix form (cf\cite{Shoukry1,Sukry008}),
\begin{equation}
\row{\sigma(t)}=A(t)\row{\sigma(0)}
\end{equation}
where $\row{S}=(\sigma_x,\sigma_y,\sigma_Z)$ and the matrix
$A=[a_{ij}]; i,j=1..3$ with coefficient $a_{ij}$ are given in the
appendix (A).

Initially, we assume that the information is encoded in the single
qubit which is prepared  in the coherent state,
\begin{equation}\label{iniQ}
\ket{\psi_q}=\cos(\theta/2)\ket{0}+e^{-i\phi}\sin(\theta/2)\ket{1},
\end{equation}
where $0\leq \phi\leq 2\pi$, $0\leq \theta \leq \pi$ and
$\ket{0},\ket{1}$ are the lower and  upper states, respectively.
The initial Bloch vector  $\row{s(0)}$ with the state (\ref{iniQ})
has the componnents,

\begin{equation}
s_x(0)=\sin\theta\cos\phi, ~~s_y(0)=\sin\theta\sin\phi,~~
s_z(0)=-\cos\theta
\end{equation}

\section{Dynamics of  information}

\subsection{Mathematical Forms}
\begin{itemize}

\item Fisher Information:

It is known that, the density operator for 2-level atomic system
is given by,
\begin{equation}
\rho_q=\frac{1}{2}(I+\row{s}\cdot\row{\sigma})
\end{equation}

where, $\row{s}=(s_x(0),s_y(0),s_z(0))$  is the  Bloch vector and
$\hat\sigma=(\hat\sigma_x,\hat\sigma_y,\hat\sigma_z)$ are the spin
Pauli operators. In terms of Bloch vector $\row{s}(\beta)$, the
quantum Fisher information(QFI)  with respect to the parameter
$\beta$ is defined as \cite{Shoukry1,Xing016},

\begin{equation}
\mathcal{F}_{\beta} = \left\{ \begin{array}{ll}
\frac{\Bigl[\row{s}(\beta)\cdot\frac{\partial{\row{s}(\beta)}}{\partial\beta}\Bigr]^2}{1-\bigl|\row{s}(\beta)\bigr|^2}
+\Bigl(\frac{\partial\row{s}(\beta)}{\partial\beta}\Bigr)^2&\row{s}(\beta)|<1,\\
\nonumber\\
\Bigl|\frac{\partial\row{s}(\beta)}{\partial\beta}\Bigr|^2 & ~|\row{s}(\beta)|=1\\
\end{array} \right.
\end{equation}
where $\beta$ is the parameter to be estimated. From Eq.(7), it is
clear that the final solution depends on the initial parameters
($\theta, \phi$) in addition to the system parameters $\delta$,
$\Omega_0$.

\item The encoded information\\
let us assume that Alice has encoded a given information to be
used in the context of quantum cryptography, for example. She will
use the Bennett and  Wiesner protocol \cite{bin}. If the final
state is given by
\begin{equation}\label{Final}
\rho(t)=\frac{1}{2}(1+s_x(t)\sigma_x+s_y(t)\sigma_y+s_z(t)\sigma_z)
\end{equation}
The amount of the coded information is given by
\begin{equation}
I_{cod}=-\lambda_1log\lambda_1-\lambda_2log\lambda_2
\end{equation}
where $\lambda_i,i=1,2$ are the eigenvalues of the state
(\ref{Final}).

\end{itemize}

\subsection{Numerical results}
 In the
following subsections, we  estimate these parameters by
calculating their corresponding QFI, $\mathcal{F}_\beta$. The
larger QFI is the higher degree of estimation for the parameter
$\beta$.

\begin{figure}[t!]
\centering
           \includegraphics[width=0.35\textwidth]{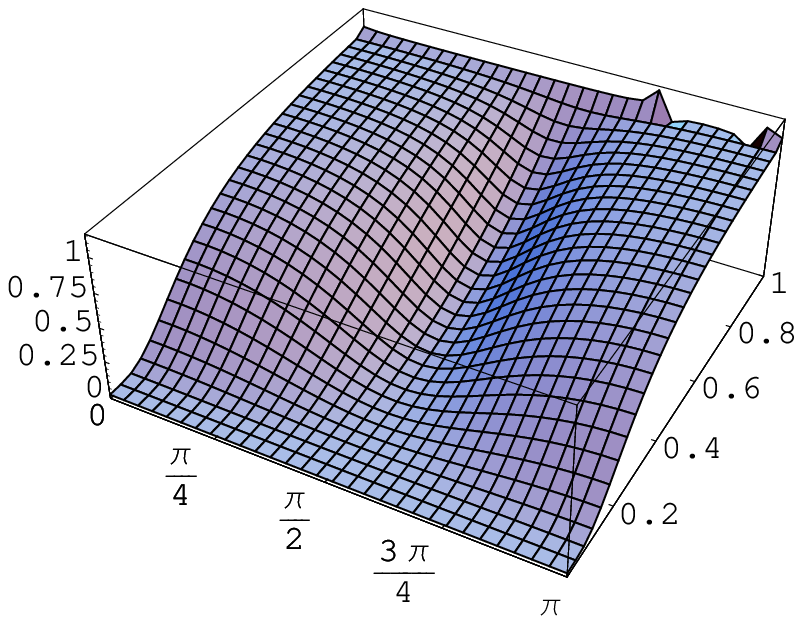}
           \put(-130,15){\large $\theta$}
     \put(-190,75){\large $\mathcal{F}_\theta$}
     \put(-15,35){\large $\Omega_0$}
     \put(-190,100){$(a)$}\\
     \vspace{0.5cm}
           \includegraphics[width=0.35\textwidth]{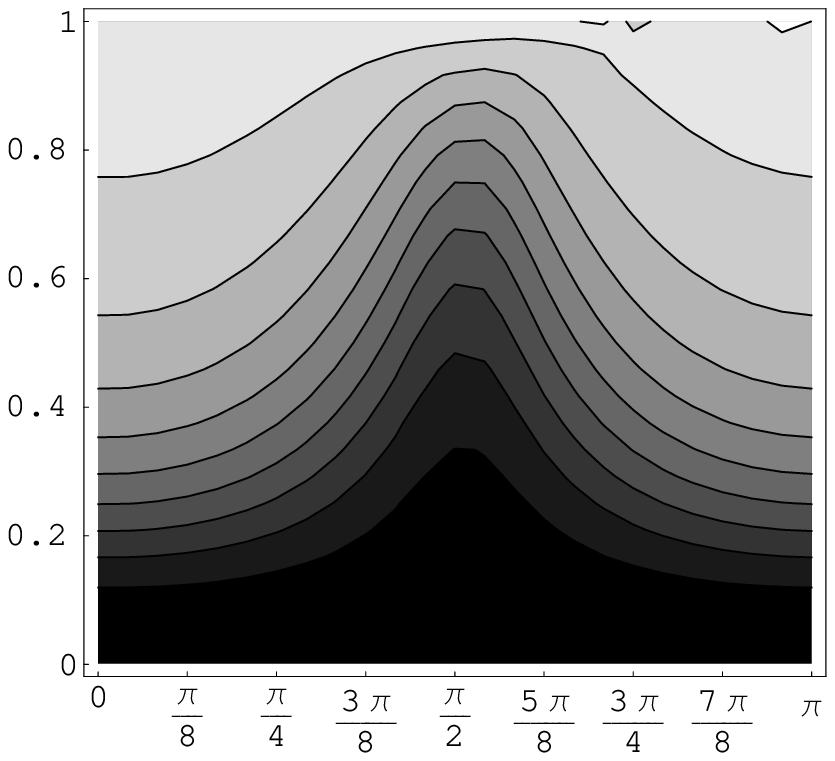}
             \put(-90,-5){\large $\theta$}
     \put(-190,95){\large $\Omega_0$}
     \put(-190,140){$(b)$}
            \caption{ (a) The pulsed Fisher information $(\mathcal{F}_\theta)$
            as a function of the frequency $\Omega_0$ and $\theta$ with $\Delta=0.2, \phi=\pi$
            (b) The contour of $\mathcal{F}_\theta$.    }
\end{figure}

Fig.(1a) describes the behavior of the quantum Fisher information
with respecte to the weight parameter $\theta$ as a function of
the frequency $\Omega_0$  at small value of the detuning
parameter, $\Delta$. It is clear that, the quantum Fisher
information $\mathcal{F}_\theta$ is almost zero for any value of
$\Omega_0<0.1$ and any initial value $\theta\in[0,\pi]$. This
means that, in this interval one can not estimate the weight
parameter.  However, for larger  values of $\Omega_0$,
$\mathcal{F}_\theta$ increases gradually to reach its maximum
values at $\Omega_0=1$.  Note also that for the  range
$0.4<\Omega_0<1$, the quantum Fisher information deceases for
$\theta\in[0,\pi/4]$ and increases for $\theta\in[\pi/4,\pi]$.
This behavior is displayed in Fig.(1b), as a contour plot, where
it is divided into different regions that have the same degree of
brightness/darkness. This means that, in these regions, the
quantum Fisher information $\mathcal{F}_\theta$ is frozen. In the
more brightened regions, the possibility of estimating the weight
parameter $\theta$ increases, while it decreases as the darkness
increases.

\begin{figure}[t!]
\centering
           \includegraphics[width=0.35\textwidth]{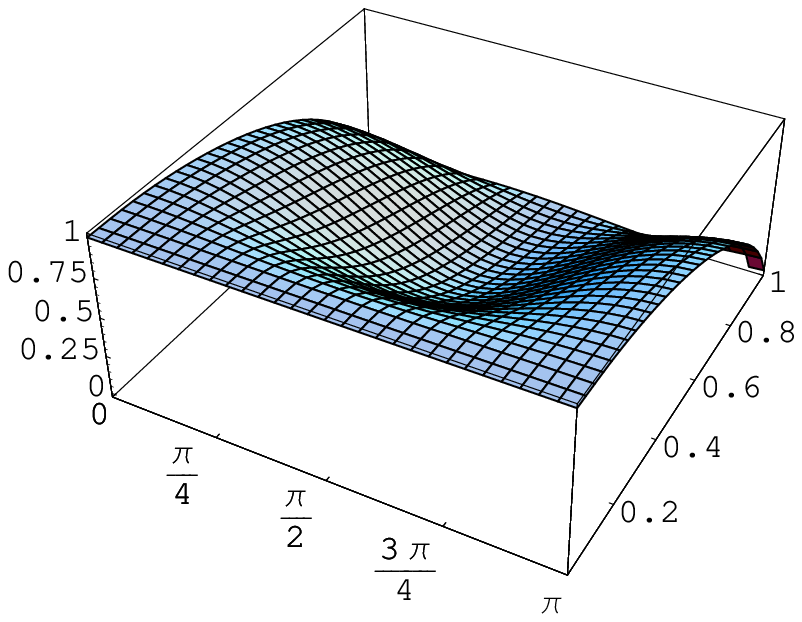}
            \put(-130,15){\large $\theta$}
     \put(-200,80){\large $I_{cod}$}
     \put(-15,35){\large $\Omega_0$}
     \put(-190,120){$(a)$}\\
     \vspace{0.5cm}
           \includegraphics[width=0.35\textwidth]{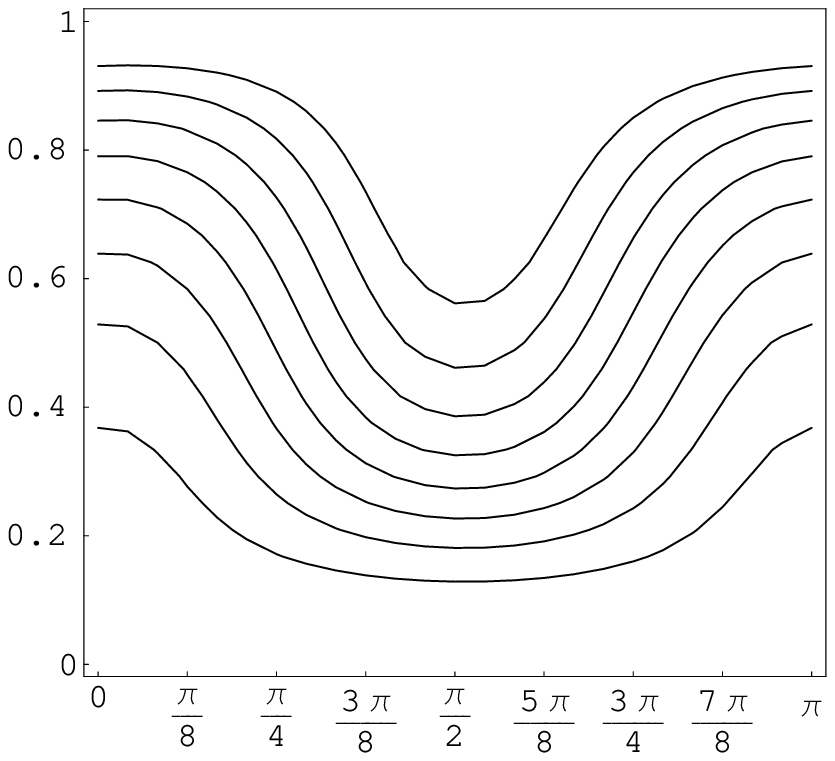}
          \put(-100,-5){\large $\theta$}
     \put(-190,95){\large $\Omega_0$}
     \put(-200,160){$(b)$}
            \caption{(a)The pulsed  encoded information, $I_{cod}$
            as a function of the frequency $\Omega_0$ and $\theta$ with $\Delta=0.2, \phi=\pi$
            (b) The contour plot of $I_{cod}$.   }
\end{figure}

In Fig.(2a), we plot the amount of the encoded information in the
pulsed state at $\Delta=0.2$. It is clear that, as soon as the
pulse is switched on, the encoded information $I_{cod}$ is maximum
at small values of $\Omega_0$ and for  any initial values of the
weight parameter, $\theta$. For larger values of $\Omega_0$, the
quantum encoded information $I_{cod}$ gradually decreases with the
minimum values of the estimation degree around $\pi=\pi/2$. The
contour plot, Fig.(2b),  displays the regions in which the encoded
information is large and  decreases as the initial weight
parameter $(\theta)$ decreases. On the other hand,  there are no
dark regions depicted which means that the encoded information
cannot vanishes.

For  larger value of the detuning parameter $(\Delta=0.9)$ the
contour of $\mathcal{F}_\theta$ in the $(\theta,\Omega_0)-$plane
is shown in , Fig.(3), where it shows the areas where the quantum
fisher information may be frozen. It is clear that, the dark
regions  are wider than those displayed for small values of the
detuning parameter (see Fig.(1b)). This means that the possibility
of estimation $\theta$ decreases as one increases $\Delta$

\begin{figure}
\centering
           \includegraphics[width=0.35\textwidth]{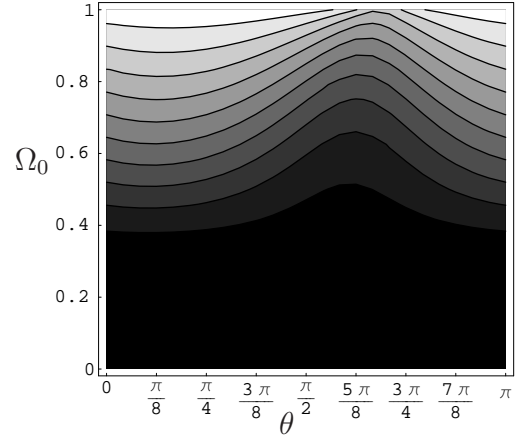}
            \put(-90,-5){\large $\theta$}
     \put(-190,95){\large $\Omega_0$}
                 \caption{ The contour plot of $\mathcal{F}_\theta$ in the
            $(\theta, \Omega_0)$-plane with $\Delta=0.9,~ \phi=\pi$ .  }
\end{figure}

\begin{figure}
\centering
              \includegraphics[width=0.35\textwidth]{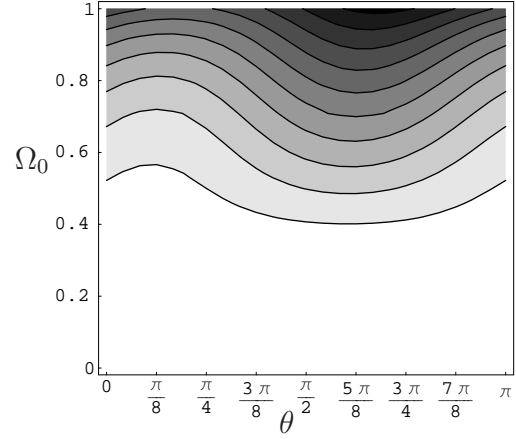}
          \put(-90,-5){\large $\theta$}
     \put(-190,95){\large $\Omega_0$}
                 \caption{The  same as Fig.(3) but for $I_{cod}$.  }
\end{figure}

The contour plot of the encoded information, $I_{cod}$ in
 Fig.(4) shows that the size of the bright
regions is much larger than that displayed in Fig.(2b). However,
the degree of brightness degreases as $\Omega_0$ increases which
means that there is a leakage of the pulsed information.

From Figs.(1-4), one may conclude that, it is possible to freeze
the coherence of the estimation degree  of   the weight parameter
$(\theta)$ by controlling  the strength of the pulse and the
detuning between the qubit and the pulse. For larger values of the
detuning and smaller values of the strength one can increase the
possibility of freezing the estimation degree of the weight
parameter.  The amount of the coded information may be maximized
as the estimation degree of the weight parameter is minimized.

\begin{figure}
\centering

           \includegraphics[width=0.35\textwidth]{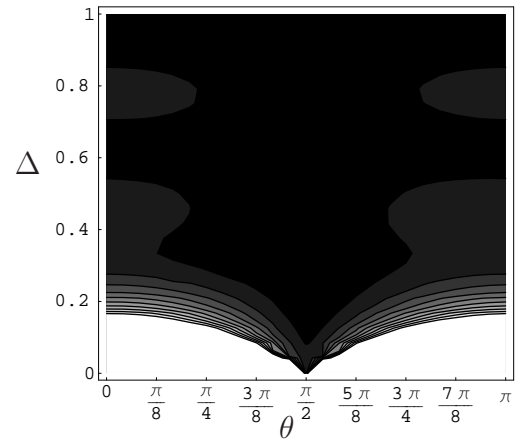}
            \put(-90,-5){\large $\theta$}
     \put(-190,95){\large $\Delta$}
               \caption{ The contour plot of $\mathcal{F}_\theta$
               in the $(\theta, \Delta)$-plane with $\Omega_0=0.1, \phi=\pi$.   }
\end{figure}

\begin{figure}
\centering

           \includegraphics[width=0.35\textwidth]{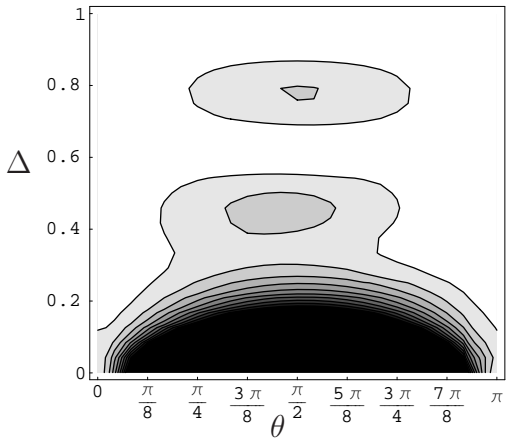}
            \put(-90,-5){\large $\theta$}
     \put(-190,95){\large $\Delta$}
    % \put(-200,155){$(b)$}
            \caption{ The same as Fig.(5) but for the encoded information $I_{cod}$.   }
\end{figure}

Figs.(5) and (6) display the  contour  behavior of the quantum
Fisher information and the encoded information, respectively, in
the $(\theta, \Delta)$-plane. It is clear that, the detuning
parameter has a decoherence effect on the Fisher information, with
a coherence effect on the encoded information.  Fig.(5) shows the
size of regions in which one may estimate the weight parameter
$(\theta)$, where the possibility of freezing the pulsed Fisher
information increases as the detuning parameter increases.

The dynamics of the pulsed encoded information, $I_{cod}$ is
depicted in Fig.(6), where it reaches its maximum values at
$\Delta=\theta=0$ and decreases suddenly as the initial weight
parameter increases and vanish completely  at
$\theta\simeq\pi/16$. However,  at any
$\theta\in[\pi/16,15\pi/16]$ and $\Delta<0.1$, the encoded
information is almost zero. For larger values of $\Delta$ and
arbitrary value of  the weight parameter, the encoded information
is almost maximum. There are two  displayed peaks where the
encoded information $I_{cod}$ is slightly decreases. Fig.(6)
represents the behavior of the encoded information in a contour
plot, where the  indicated dark regions  are very small, while the
brightened  regions are large and reach its maximum values at
large values of the detuning parameter.

Further, it is clear that,  one can maximize the amount of pulsed
encoded information at the expense of minimizing the  estimation
degree of the weight parameter $(\theta)$. This phenomena may be
achieved by decreasing the pulse strength and increasing the
detuning between the qubit and the pulse.

\begin{figure}
\centering
               \includegraphics[width=0.35\textwidth]{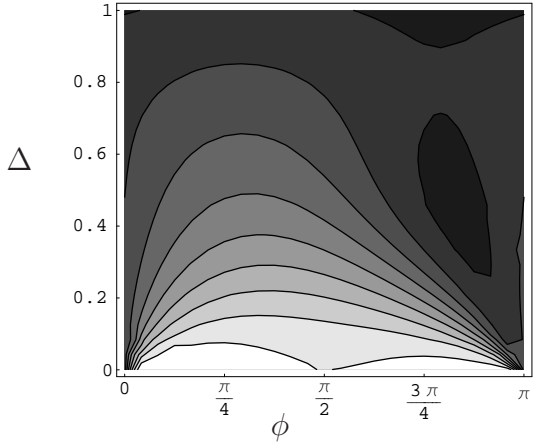}
            \put(-100,-5){\large $\phi$}
     \put(-200,95){\large $\Delta$}
                \caption{ The contour plot of  $\mathcal{F}_\phi$  in the $(\phi,\Delta)$-plane
            with   $\Omega_0=0.5, \theta=\pi$.   }
\end{figure}

\begin{figure}
\centering
            \includegraphics[width=0.35\textwidth]{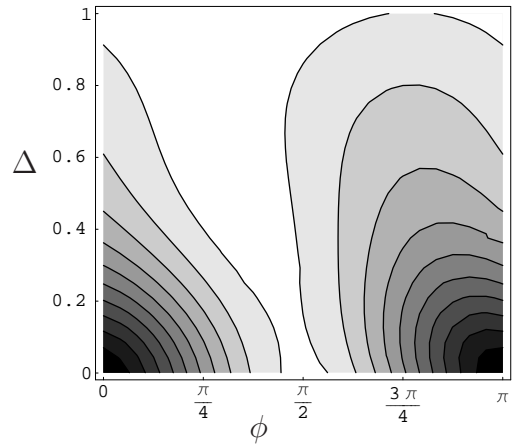}
            \put(-100,-5){\large $\phi$}
     \put(-190,95){\large $\Delta$}
    % \put(-190,160){$(b)$}
            \caption{ The   same as Fig.(7) but for the encoded information $I_{cod}$.   }
\end{figure}

In Figs.(7) and (8), we investigate the behavior of the Fisher
information $\mathcal{F}_\phi$  and the  encoded information when
the phase parameter $(\phi)$ is estimated,  such that the driven
qubit is initially prepared in the state $e^{-i\phi}\ket{1}$,
namely,  we set the weight parameter $\theta=\pi$.  It is clear
that, the larger values of the detuning has a decoherence effect
on the Fisher information,$\mathcal{F}_\phi $, where it decreases
as $\Delta$ increases.  Fig.(7) displays the  area in which the
Fisher information is frozen, where the degree of the darkness
indicates the estimation. As $\Delta$ increases, the darkness
increases which means that, the possibility of estimating the
phase parameter $(\phi)$ decreases. On the other hand, Fig.(8),
for the encoded information shows that the brightness increases as
$\Delta$ increases and the maximum bounds are displayed around
$\phi=\pi/2$.

\begin{figure}
\centering
             \includegraphics[width=0.35\textwidth]{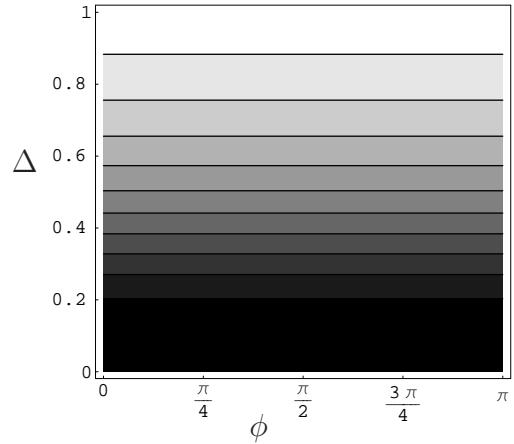}
            \put(-100,-5){\large $\phi$}
     \put(-190,95){\large $\Delta$}
                 \caption{ The  encoded information $I_{cod}$, in
                 the $(\phi,\Delta)$-plane  with  $\Omega_0=0.5, \theta=0$.   }
\end{figure}

Fig.(9) describes the contour  behavior of the encoded information
$I_{cod}$  for a different initial state setting, where it is
assumed that the qubit is initially prepared in the state,
$\ket{\psi(0)}=\ket{0}$, namely, $\theta=0$. This means that, the
initial state doesn't depend on the phase $\phi$ and may be taken
arbitrary. On the other hand, the freezing phenomena of the pulsed
encoded information, $I_{cod}$ is depicted at small values of the
detuning and the degree of freezing decreases as the detuning
increases.

\section{CONCLUSIONS}

In this contribution, we investigate the relation between the
pulsed  Fisher information of the qubit's parameters and the
encoded information. The suggested system consists of a single
qubit driven by a rectangular pulse. These physical quantities,
the Fisher and the encoded information, are discussed for
different values of the pulse strength and the detuning between
the qubit and the pulse.

In case of estimating the weight parameter $(\theta)$, it is shown
that, large values of the  pulse strength increase the possibility
of estimating the weight parameter and  decreases the capacity of
encoded information in the qubit. Large values of the detuning
increase the size of the  frozen areas for the   two  physical
quantities; estimation degree and the channel capacity. However,
for  increased detuning, the estimation degree of the weight
parameter increases, while the channel capacity decreases. The
behavior of the Fisher information and the encoded information as
functions of the detuning parameter is discussed for  small values
of the pulse strength. It is shown that, it is possible to
maximize the channel capacity at the expense of the estimation
degree. Moreover,  one can always freeze both quantities for any
initial state setting of the weight parameter.

The behavior of Fisher information and the coded information is
discussed when  the phase parameter $(\phi)$ is estimated. In this
case, the initial phase plays an important role on the
decoherence/ coherence effect of the pulse.  The results, show
that the encoded information doesn't depend on $\phi$, while the
Fisher information depend on it.

{\it In conclusion}, it is possible to  freeze the Fisher
information  and the amount of the encoded information for both
qubit parameters $(\theta,\phi)$.  One can increase the  size of
the frozen area of the encoded information at the expense  of
Fisher information. We show that, the encoded information doesn't
depend on the  phase ($\phi$). We expect that, these results may
be useful  in the context of cryptography and secure
communications.

% \addtolength{\textheight}{-12cm}   % This

%%%%%%%%%%%%%%%%%%%%%%%%%%%%%%%%%%%%%%%%%%%%%%%%%%%%%%%%%%%%%%%%%%%%%%%%%%%%%%%%

%%%%%%%%%%%%%%%%%%%%%%%%%%%%%%%%%%%%%%%%%%%%%%%%%%%%%%%%%%%%%%%%%%%%%%%%%%%%%%%%

%%%%%%%%%%%%%%%%%%%%%%%%%%%%%%%%%%%%%%%%%%%%%%%%%%%%%%%%%%%%%%%%%%%%%%%%%%%%%%%%
%\section*{ACKNOWLEDGMENT}

%%%%%%%%%%%%%%%%%%%%%%%%%%%%%%%%%%%%%%%%%%%%%%%%%%%%%%%%%%%%%%%%%%%%%%%%%%%%%%%%

\section{Appendix(A)}
The coefficients $a_{ij}$ of the matrix $A(t)$ in Eq.(4)are as
follows
\begin{eqnarray*}
a_{11}&=&\frac{1}{\eta}+\delta^2\cos(\tau\sqrt{\eta})-\delta\lambda_1
\nonumber\\
a_{12}&=&\frac{1}{2}(1+\frac{\lambda_2}{\eta}+\delta\lambda_1
\nonumber\\
a_{13}&=&\frac{\delta}{\eta}\lambda_3+\lambda_1
\nonumber\\
a_{21}&=&\frac{\lambda_4}{2\eta}+\lambda_1
\nonumber\\
a_{22}&=&\cos(\tau\sqrt{\eta})-\delta\lambda_1
\nonumber\\
a_{23}&=&\frac{\delta}{\eta}\lambda_3-\delta\lambda_1
\nonumber\\
a_{31}&=&\frac{\delta}{\eta}\lambda_3, \quad
a_{32}=\lambda_{1},\quad a_{33}=\frac{\lambda_2}{\eta}
\end{eqnarray*}
where,
\begin{eqnarray*}
\lambda_1&=&\frac{1}{\sqrt{\eta}}\sin(\tau\sqrt{\eta}) \nonumber\\
\lambda_2&=&\delta^2+\cos(\tau\sqrt{\eta}),, \nonumber\\
\lambda_3&=&\frac{1}{2}\eta\lambda_1^2, \quad
\lambda_4=1+(\eta+\delta^2)\cos(\tau\sqrt{\eta})
\end{eqnarray*}
and $\delta=\frac{\Delta}{\Omega_0}, \eta=1+\delta^2,\eta=\Omega_0
t$
\end{document}